\documentclass{PoS}
\usepackage{amsmath}
\usepackage{cancel}

\newcommand{\be}{\ensuremath{\beta} }
\newcommand{\De}{\ensuremath{\Delta} }
\newcommand{\la}{\ensuremath{\lambda} }
\newcommand{\chibar}{\ensuremath{\overline\chi} }
\newcommand{\psibar}{\ensuremath{\overline\psi} }
\newcommand{\vev}[1]{\ensuremath{\left\langle #1 \right\rangle} }
\newcommand{\pbp}{\ensuremath{\vev{\psibar\psi}} }
\newcommand{\X}{\ensuremath{\!\times\!} }
\newcommand{\Sb}{\ensuremath{\cancel{S^4}} }
\newcommand{\fig}[1]{Fig.~\ref{#1}}
\newcommand{\refcite}[1]{Ref.~\cite{#1}}
\newcommand{\secref}[1]{Section~\ref{#1}}
\newcommand{\mysection}[1]{\vspace{-12 pt}\section{#1}\vspace{-6 pt}}
\newcommand{\mutesection}[1]{\vspace{-12 pt}\section*{#1}\vspace{-6 pt}}
\newcommand{\mysubsection}[1]{\vspace{-12 pt}\subsection{#1}\vspace{-6 pt}}

\title{Bulk and finite-temperature transitions \\ in SU(3) gauge theories with many light fermions}
\ShortTitle{Bulk and finite-temperature transitions with many light fermions}

\author{\speaker{David Schaich}, Anqi Cheng, Anna Hasenfratz and Gregory Petropoulos \\
  Department of Physics, University of Colorado, Boulder, CO 80309 \\
  Email: \email{schaich@pizero.colorado.edu}
}

\abstract{ 
  We investigate finite-temperature transitions in SU(3) lattice gauge theories with $N_f = 8$ and 12 staggered fermions in the fundamental representation.
  For both of these systems, we have observed a strongly-coupled lattice phase in which the single-site shift symmetry of the staggered action is spontaneously broken.
  Here we report new results for finite-temperature transitions on $24^3\X12$ and $32^3\X16$ lattice volumes, contrasting the 8- and 12-flavor systems.
  While the $N_f = 12$ finite-temperature transitions accumulate at the bulk transition bounding the strongly-coupled lattice phase, the $N_f = 8$ finite-temperature transitions are able to pass through the bulk transition, and behave as expected for a QCD-like system.
  We discuss our current results and the work in progress to complete our investigation of the finite-temperature phase diagram.
}

\FullConference{30th International Symposium on Lattice Field Theory \\ 24--29 June 2012 \\ Cairns, Australia}

\begin{document}
\mysection{Introduction} 
We recently reported the observation and characterization of a novel phase in SU(3) lattice gauge theories with $N_f = 8$ and 12 nHYP-smeared staggered fermions~\cite{Cheng:2011ic}.
From a variety of observables (including the meson spectrum, static potential, low-lying eigenvalues of the massless staggered Dirac operator, renormalization group blocked plaquette and Polyakov loop, and newly-developed order parameters), we established that the single-site shift symmetry (``$S^4$'') of the staggered action is spontaneously broken (``$\Sb$'') in this phase.
In terms of continuum symmetries, the \Sb phase possesses both chiral symmetry and axial U(1)$_A$ symmetry in the chiral limit, even though the Polyakov loop and static potential clearly indicate confinement.
However, we argued that the \Sb phase is likely to be a purely lattice phase with no continuum limit, on the grounds that \vspace{-6 pt}
\begin{enumerate}
  \item its combination of confinement and chiral symmetry is forbidden by the continuum 't~Hooft anomaly matching condition; \vspace{-6 pt}
  \item it is bounded by first-order bulk (zero-temperature) phase transitions; and \vspace{-6 pt}
  \item it appears in both 8- and 12-flavor systems, which we believe exhibit different infrared dynamics (QCD-like spontaneous chiral symmetry breaking and IR conformality, respectively).
\end{enumerate}
\vspace{-6 pt}
Similar observations were also reported (with somewhat different interpretations) by other groups exploring the 12-flavor system with different staggered actions~\cite{Deuzeman:2011pa, Schroeder:2011LAT, Nunez:2012LAT}.

\begin{figure}[ht]
  \centering
  \includegraphics[width=0.45\linewidth]{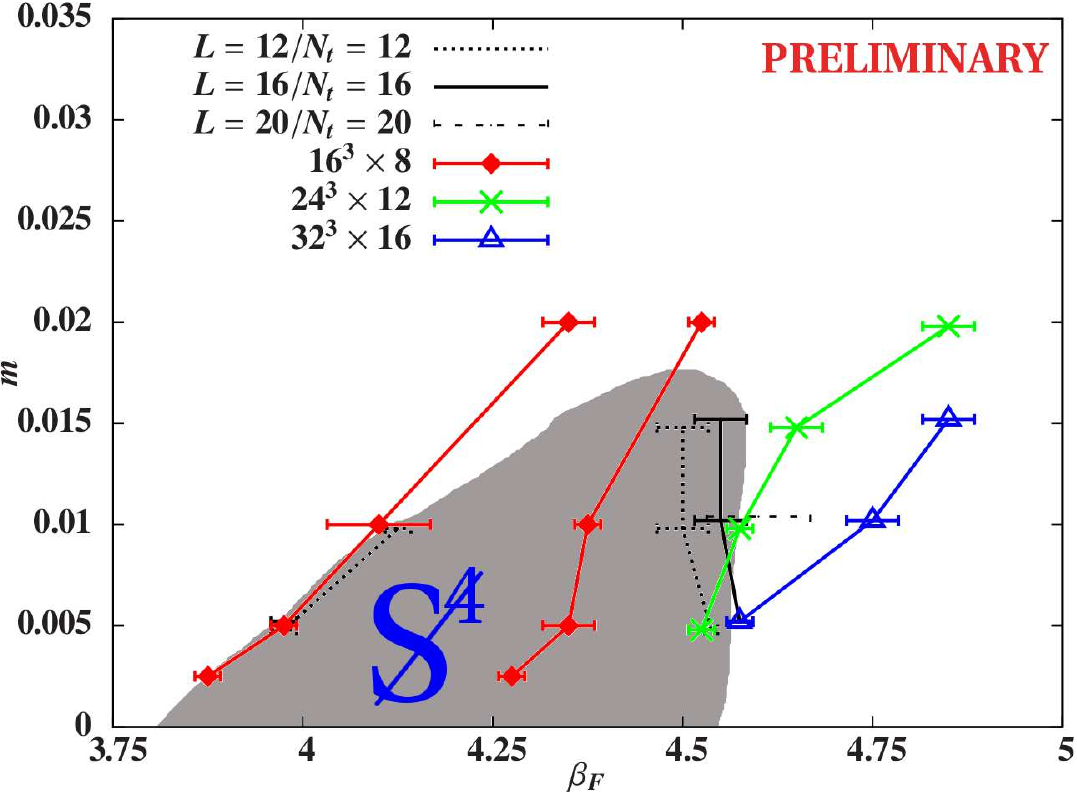}\hfill\includegraphics[width=0.45\linewidth]{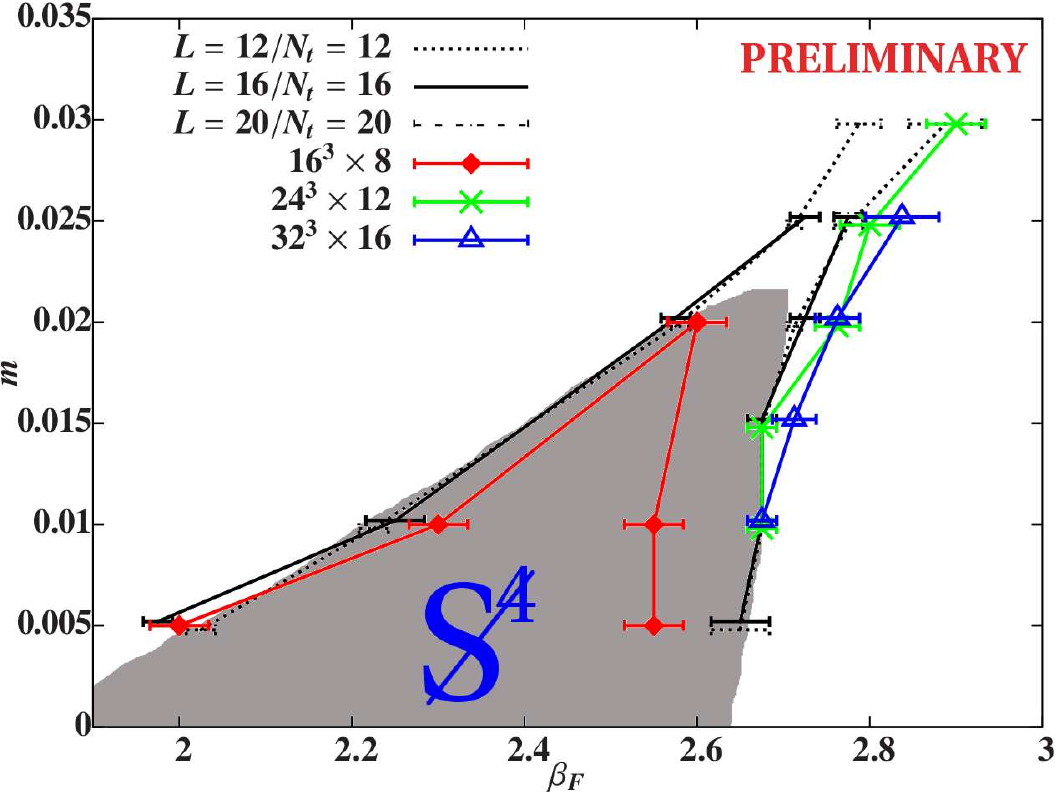}
  \caption{\label{fig:phases} Phase diagrams in the $\be_F$--$m$ plane for $N_f = 8$ (left) and 12 (right).  The \Sb phase is shaded and the bulk transitions bounding it are shown in black (including preliminary $40^3\X20$ results).  The colored points indicate finite-temperature transitions determined primarily from the RG-blocked Polyakov loop and eigenvalue density.  (The red diamonds at stronger couplings are determined from $\pbp$.)}
\end{figure}

We have since extended our investigations of the 8- and 12-flavor systems through several complementary analyses, two of which (studies of the Dirac eigenmode scaling and Monte Carlo renormalization group) are discussed in other contributions to these proceedings~\cite{Hasenfratz:2012LAT, Petropoulos:2012LAT}.
In this work we consider finite-temperature transitions in the 8- and 12-flavor systems, focusing on their behavior around the bulk transition that separates the \Sb phase from the weak-coupling phase connected to the continuum limit.
Working primarily with lattice volumes $L^3\X N_t = 24^3\X12$ and $32^3\X16$, we observe an interesting contrast between the 8- and 12-flavor systems, shown in \fig{fig:phases}.
For $N_f = 12$, the finite-temperature transitions for $N_t = 12$ and 16 are indistinguishable, and fall on top of the bulk transition bounding the \Sb phase.
For $N_f = 8$, the finite-temperature transitions can pass through the bulk transition, moving to weaker couplings (larger $\be_F$) as $N_t$ increases, in agreement with RG scaling.
However, the 8-flavor system behaves differently at the lightest mass we consider, $m = 0.005$, than at $m \geq 0.01$.
We discuss this situation in \secref{sec:disc}.

In \secref{sec:obs} we present the observables that play the most prominent role in establishing the results shown in \fig{fig:phases}.
First, the \Sb order parameters we introduced in \refcite{Cheng:2011ic} are the most robust means to resolve the bulk transitions around the \Sb phase, for $T > 0$ as well as zero-temperature systems.
Next, we review the validity of our procedure to improve the signal in the Polyakov loop by measuring it on RG-blocked lattices.
Finally, the Dirac eigenvalue density $\rho(\la)$ is sensitive to the chiral properties of the system, providing additional information complementary to that from the other observables.
In addition to distinguishing between chirally broken and chirally symmetric systems, we can also use $\rho(\la)$ to identify the \Sb phase.
We conclude in \secref{sec:disc} by discussing the implications of our current results, and our plans for completing this investigation.

\mysection{\label{sec:obs} Observables for bulk and finite-temperature transitions} 
\vspace{6 pt}
\mysubsection{\Sb order parameters} 
The staggered action possesses an exact symmetry under the single-site shift transformation
\begin{align}
  \chi(n) & \to \xi_{\mu}(n) \chi(n + \mu), & \chibar(n) & \to \xi_{\mu}(n) \chibar(n + \mu), & U_{\mu}(n) & \to U_{\mu}(n + \mu), \label{eq:shift_sym}
\end{align}
where $\xi_\mu(n) = (-1)^{\sum_{\nu > \mu} n_\nu}$. 
Two order parameters of this symmetry are
\begin{align}
  \Delta P_{\mu} = & \vev{\mbox{ReTr } \square_{n, \mu} - \mbox{ReTr } \square_{n + \mu, \mu}}_{n_\mu \ {\rm even}} \\
  \Delta U_{\mu} = & \vev{\alpha_{\mu}(n) \chibar(n) U_{\mu}(n) \chi(n + \mu) - \alpha_{\mu}(n + \mu)\chibar(n + \mu) U_{\mu}(n + \mu) \chi(n + 2\mu)}_{n_\mu {\rm \ even}}.
\end{align}
In these expressions $\square_{n, \mu}$ indicates the plaquettes originating at lattice site $n$ that include links in the $\mu$ direction; $U$ and $\chi$ are the gauge and fermion fields, respectively; $\alpha_{\mu}(n) = (-1)^{\sum_{\nu < \mu} n_{\nu}}$; and the expectation value $\vev{\cdots}_{n_{\mu} \ {\rm even}}$ is taken only over sites whose $\mu$ component is even.

In \refcite{Cheng:2011ic} we showed that $\Delta P_{\mu}$ and $\Delta U_{\mu}$ vanish in both the weak-coupling phase and the familiar chirally-broken lattice phase at strong coupling.
Nonzero expectation values (in one or more directions $\mu$) characterize the intermediate \Sb phase where the single-site shift symmetry is spontaneously broken.
Even though the \Sb phase is bounded by bulk (zero-temperature) phase transitions, we observe the same transitions in the \Sb order parameters on the finite-temperature lattices we consider here.
That is, $24^3\X12$ calculations show the same transition in $\Delta P_{\mu}$ and $\Delta U_{\mu}$ as do $12^3\X24$ systems, and similarly for $32^3\X16$ compared to $16^3\X32$; these results are combined in \fig{fig:phases}, labelled ``$L = 12 / N_t = 12$'', etc.
Our preliminary $40^3\X20$ data reveal only the \Sb transition, not yet the finite-temperature transition.

\begin{figure}[ht]
  \centering
  \includegraphics[width=0.45\linewidth]{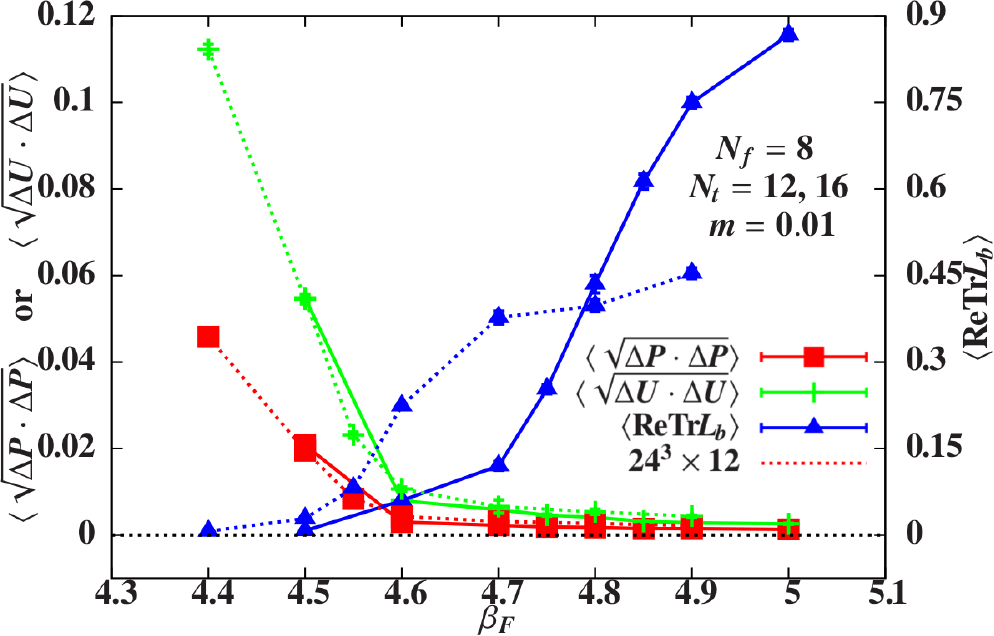}\hfill\includegraphics[width=0.45\linewidth]{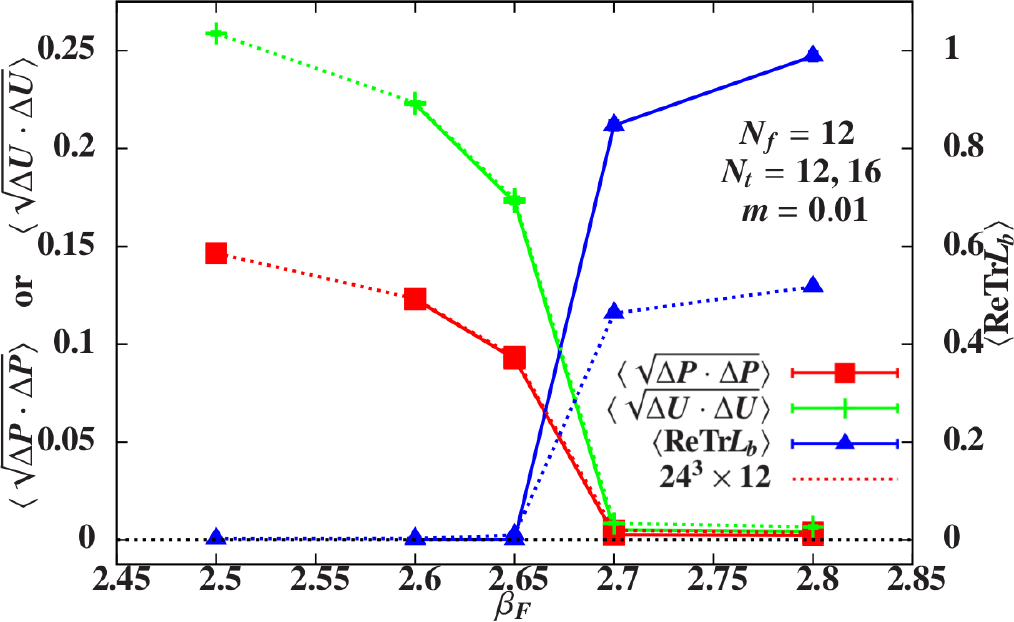}
  \caption{\label{fig:trans} Signals of transitions in $\be_F$ with fixed $m = 0.01$ for $N_f = 8$ (left) and 12 (right).  The solid lines correspond to $32^3\X16$ volumes, dotted lines to $24^3\X12$.  The green crosses and red squares are derived from the \Sb order parameters sensitive to the bulk transition bounding the \Sb phase.  The blue triangles are RG-blocked Polyakov loop results, one indicator of the finite-temperature deconfinement transition.}
\end{figure}

In \fig{fig:trans} we show the euclidean norms $\sqrt{\Delta P\cdot\Delta P}$ and $\sqrt{\Delta U\cdot\Delta U}$ on $24^3\X12$ and $32^3\X16$ volumes for both the 8- and 12-flavor systems at fixed fermion mass $m = 0.01$.
While these euclidean norms are not themselves order parameters, they are sensitive to the \Sb transition because they can only be large if $\Delta P_{\mu}$ and $\Delta U_{\mu}$ develop nonzero expectation values in one or more directions.
The \Sb transition is volume-independent, with $24^3\X12$ and $32^3\X16$ results on top of each other.

\mysubsection{RG-blocked Polyakov loop} 
\fig{fig:trans} also includes RG-blocked Polyakov loop results, which show the $N_f = 8$ finite-temperature deconfinement transitions moving as $N_t$ increases, while those for $N_f = 12$ do not.
In pure-gauge systems, the Polyakov loop $\vev{\mbox{Tr}L}$ is an order parameter, and it remains sensitive to deconfinement even in the presence of dynamical fermions.
However, $\vev{\mbox{Tr}L}$ becomes small and noisy as $N_t$ increases, which motivated us to construct the improved observable $\vev{\mbox{Tr}L_b}$ by measuring the Polyakov loop on RG-blocked lattices.
$\vev{\mbox{Tr}L_b}$ has the usual $Z_3$ symmetry in the pure-gauge theory, and can also be thought of as an extended observable on the original, unblocked lattices.

\begin{figure}[ht]
  \centering
  \includegraphics[width=0.45\linewidth]{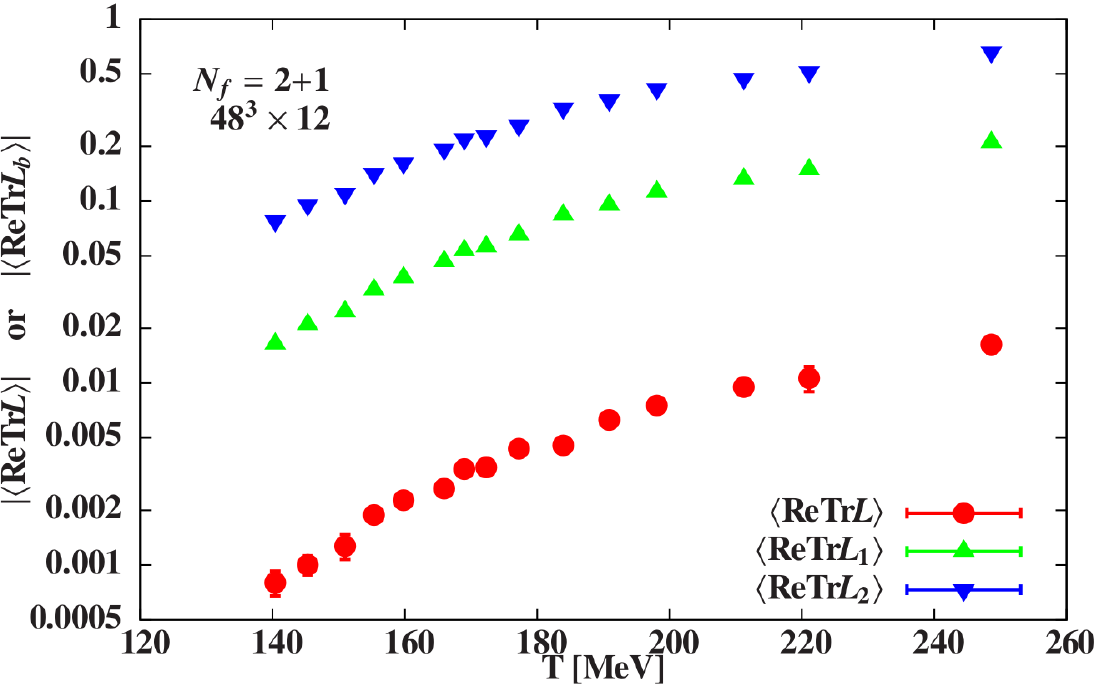}\hfill\includegraphics[width=0.45\linewidth]{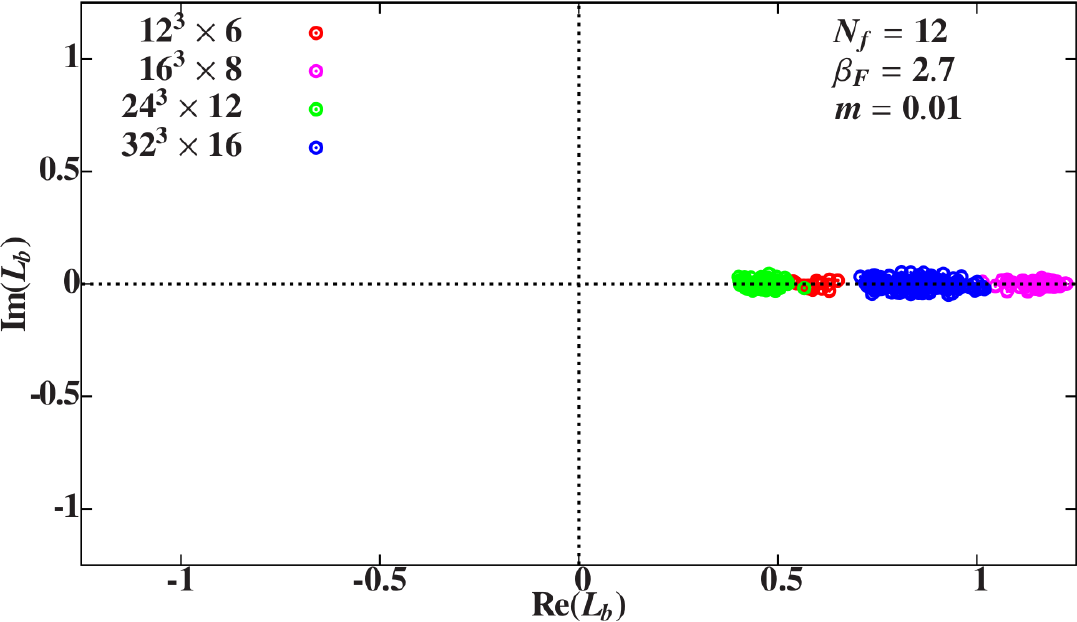}
  \caption{\label{fig:poly} Left: RG-blocked Polyakov loop results from $N_f = 2$+1 ensembles generated by the HotQCD Collaboration~\protect\cite{Bazavov:2011nk}, on a log scale.  The red circles are the unblocked observable, green triangles are once-blocked (i.e., smeared), and blue inverted triangles are twice-blocked.  Right: Maximally-blocked Polyakov loop data for the $N_f = 12$ system at $\be_F = 2.7$ and $m = 0.01$ on volumes $12^3\X6$ through $32^3\X16$.}
\end{figure}

Our RG-blocking transformation consists of two HYP smearings with $\alpha = (0.6, 0.2, 0.2)$, as discussed in \refcite{Hasenfratz:2010fi}.
The once-blocked $\vev{\mbox{Tr}L_1}$ is therefore equivalent to the usual Polyakov loop measured with smeared links.
As the number of blocking steps increases, the effective $N_t$ of the lattice is repeatedly halved, amplifying the signal in $\vev{\mbox{Tr}L_b}$.
To check that the RG-blocked Polyakov loop behaves appropriately, we measured it on $N_f = 2$+1 configurations made available by the HotQCD Collaboration~\cite{Bazavov:2011nk}.
The results in the left panel of \fig{fig:poly} show that the RG-blocked Polyakov loop captures the same physics as the familiar observable, the only change being order-of-magnitude enhancements in the signal.
At the QCD transition temperature $T_c \approx 155$ MeV, the maximally-blocked Polyakov loop on these $48^3\X12$ lattices is $\vev{\mbox{ReTr}L_2} = 0.142(1)$, while the unblocked value is $\vev{\mbox{ReTr}L} = 0.00188(13)$.

The right panel of \fig{fig:poly} shows maximally-blocked Polyakov loop data for the $N_f = 12$ system at $\be_F = 2.7$ and $m = 0.01$ on volumes $12^3\X6$ through $32^3\X16$.
Even though the unblocked Polyakov loop on these lattices is indistinguishable from zero for $N_t \gtrsim 12$, the RG-blocked data indicate that the system is deconfined through at least $N_t = 16$.
It is important to note that even the maximally-blocked Polyakov loop can still indicate confinement: this is the case for the $N_f = 8$ system with $N_t = 16$ at $\be_F = 4.7$ and $m = 0.01$.
\fig{fig:trans} shows that this point is on the weak-coupling side of the \Sb phase, but not yet deconfined.

\begin{figure}[ht]
  \centering
  \includegraphics[width=0.45\linewidth]{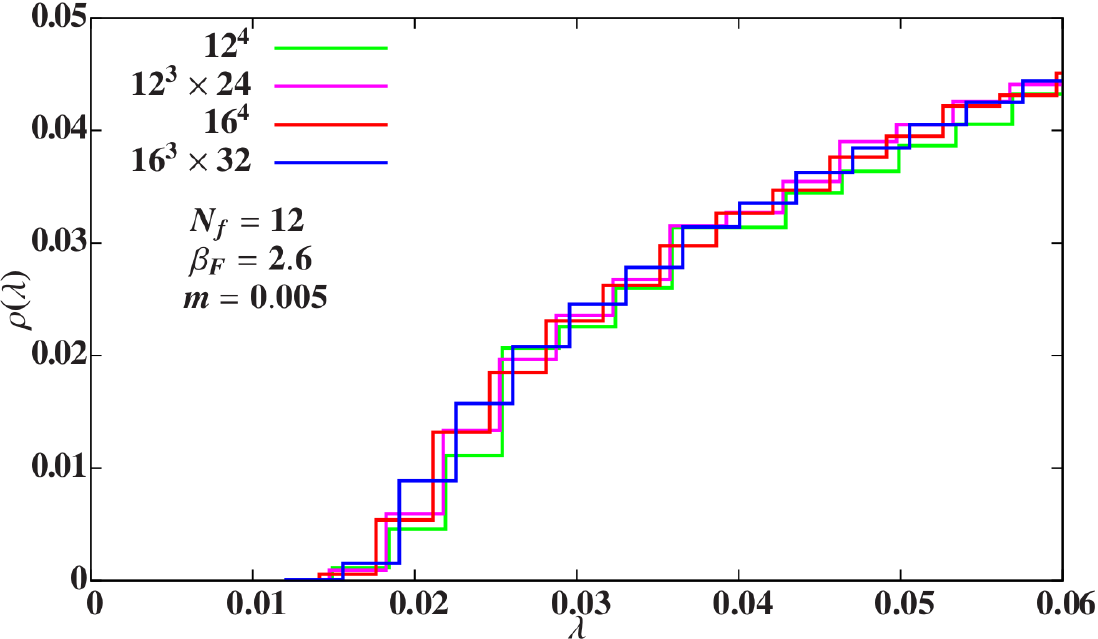}
  \caption{\label{fig:rho_edge} The $N_f = 12$ eigenvalue density $\rho(\la)$ determined from the 200 lowest-lying eigenvalues on each of four volumes in the \Sb phase ($\be_F = 2.6$, $m = 0.005$).}
\end{figure}

\mysubsection{Eigenvalue densities} 
While investigating the 12-flavor \Sb phase in \refcite{Cheng:2011ic}, we found the low-lying eigenvalues of the massless staggered Dirac operator to be a particularly striking observable.
In the \Sb phase we observed a soft edge (a gap in the infinite-volume extrapolation of the eigenvalue density $\rho(\la)$), indicating restoration of both chiral symmetry and axial U(1)$_A$ symmetry in the chiral limit.
\fig{fig:rho_edge} shows nearly indistinguishable $\rho(\la)$ on four zero-temperature volumes in the \Sb phase, which extrapolate to the soft edge $\la_0 = 0.0175(5)$.
While a soft edge is unusual, $\rho(0) = 0$ also for chirally symmetric systems at high temperatures above the finite-temperature transition.
Chiral symmetry breaking is characterized by $\rho(0) \ne 0$.

\begin{figure}[ht]
  \centering
  \includegraphics[width=0.45\linewidth]{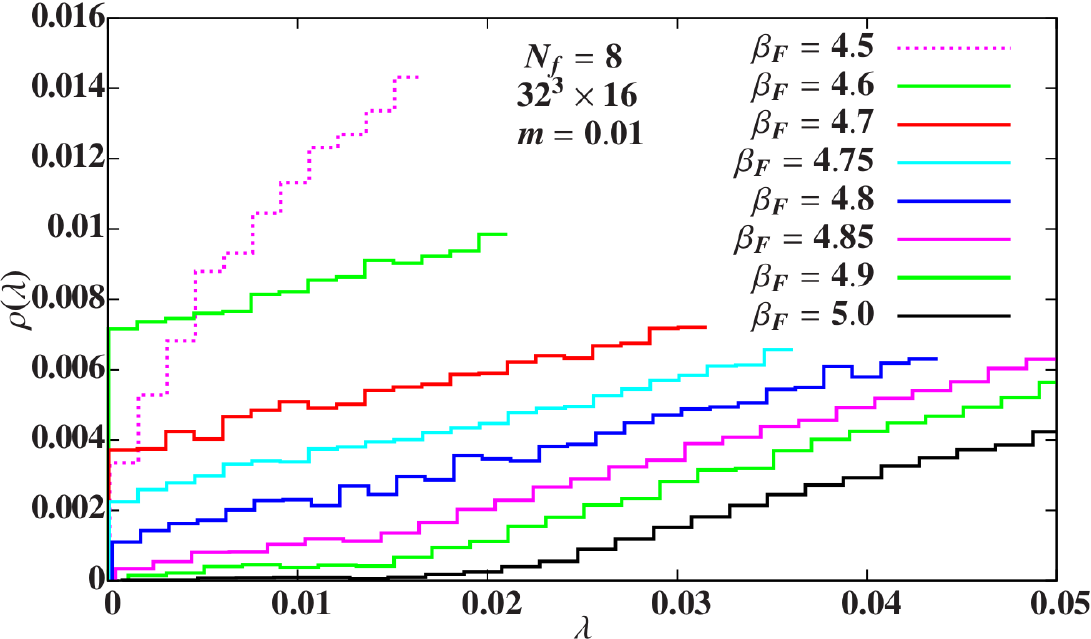}\hfill\includegraphics[width=0.45\linewidth]{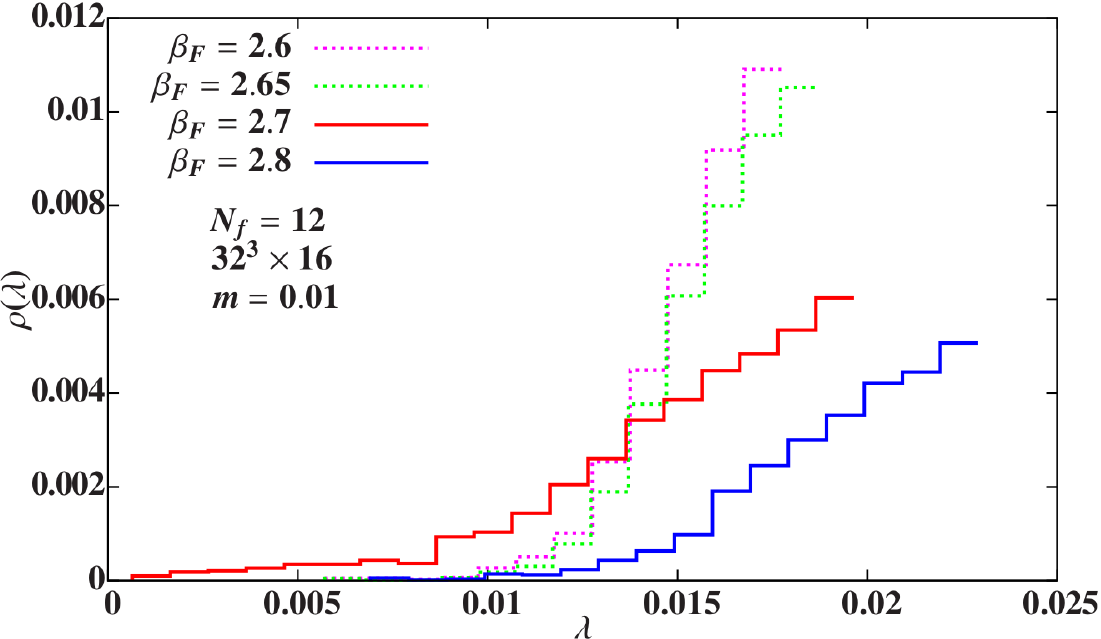}
  \caption{\label{fig:rho} Eigenvalue density $\rho(\la)$ for fixed $m = 0.01$ on $32^3\X16$ volumes for $N_f = 8$ (left) and $N_f = 12$ (right).  The dotted lines indicate $\be_F$ in the \Sb phase.}
\end{figure}

The eigenvalue density $\rho(\la)$ therefore provides information about both the bulk and finite-temperature transitions.
This is illustrated in \fig{fig:rho}, which presents results for the 8- and 12-flavor systems with $m = 0.01$ on $32^3\X16$ volumes.
In the \Sb phase (dotted lines), $\rho(\la)$ shows a steep slope for both systems, and a clear gap for $N_f = 12$.
As the coupling becomes weaker, $N_f = 8$ moves from the \Sb phase to a chirally broken system with $\rho(0) \ne 0$; the transition to a chirally symmetric system only occurs at even weaker couplings around $\be_F \approx 4.85$.
For $N_f = 12$, in contrast, we move directly from the \Sb phase to a chirally symmetric system.
In both cases, $\rho(\la)$ agrees with the behavior of the \Sb order parameters and RG-blocked Polyakov loop shown in \fig{fig:trans}.

\mysection{\label{sec:disc} Discussion} 
The selected observables reviewed in the previous section are those that we find most useful for identifying both the \Sb bulk transition (order parameters and eigenvalues) as well as the finite-temperature transitions (RG-blocked Polyakov loop and eigenvalues).
The results of these investigations are shown in \fig{fig:phases}, which we now discuss in more detail.

The most striking feature of \fig{fig:phases} is the contrast between $N_f = 8$ (for which the finite-temperature transitions can pass through the bulk transition) and $N_f = 12$ (for which the finite-temperature transitions accumulate at the bulk transition).
This behavior is clearly visible in both of Figs.~\ref{fig:trans} and \ref{fig:rho} for the case of $m = 0.01$.
\fig{fig:trans} shows the \Sb order parameters falling to zero at the same $\be_F$ on $24^3\X12$ and $32^3\X16$ volumes.
For $N_f = 12$ the RG-blocked Polyakov loop also jumps to a large value at this $\be_F$, while for $N_f = 8$ $\vev{\mbox{Tr}L_b}$ only rises at weaker couplings.
Similarly, in \fig{fig:rho} we observe $\rho(0) \ne 0$ (chiral symmetry breaking) for 8-flavor systems on the weak-coupling side of the \Sb phase.
The 12-flavor system on volumes up to $32^3\X16$ is deconfined and chirally symmetric ($\rho(0) = 0$) for all couplings weaker than the \Sb phase.

A puzzling feature of \fig{fig:phases} is the different behavior of the 8-flavor system at $m = 0.005$ compared to larger masses.
For $m \geq 0.01$, the change in $\be_F$ between the $N_t = 12$ and 16 transitions is $\De\be_F \approx 0.2$, in rough agreement with the two-loop renormalization group prediction $\De\be_F \approx 0.25$.
At $m = 0.005$ the shift is much smaller ($\De\be_F \approx 0.05$), and the transition in the \Sb order parameters moves the same amount instead of being volume-independent as at $m \geq 0.01$.

\begin{figure}[ht]
  \centering
  \includegraphics[width=0.45\linewidth]{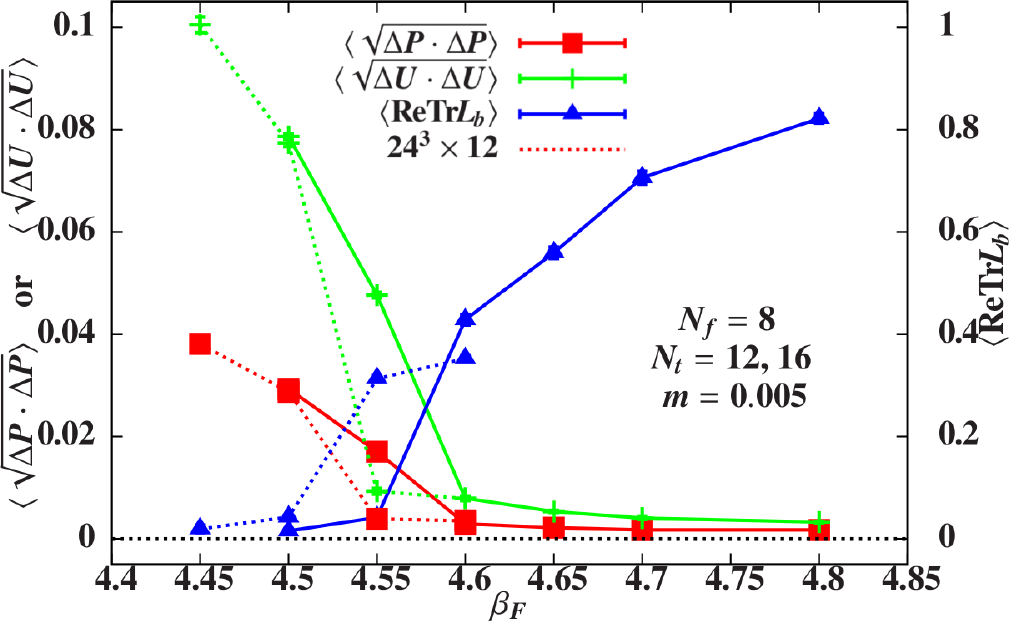}\hfill\includegraphics[width=0.45\linewidth]{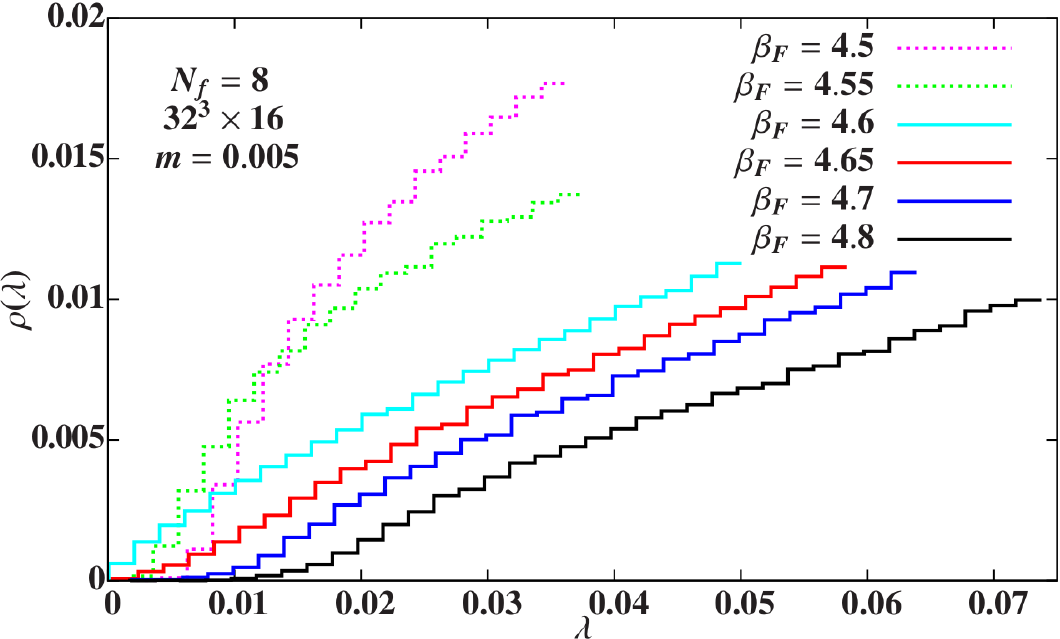}
  \caption{\label{fig:m005} $N_f = 8$ results for $m = 0.005$.  Left: \Sb order parameter and RG-blocked Polyakov loop results as in Fig.~\protect\ref{fig:trans}.  Right: eigenvalue density $\rho(\la)$ as in Fig.~\protect\ref{fig:rho}.}
\end{figure}

In \fig{fig:m005} we show the 8-flavor results at $m = 0.005$ for the \Sb order parameters, RG-blocked Polyakov loop and eigenvalue density, corresponding to that in Figs.~\ref{fig:trans} and \ref{fig:rho} for $m = 0.01$.
In contrast to both the 8- and 12-flavor results in \fig{fig:trans}, the transition in the \Sb order parameters for $N_f = 8$ at $m = 0.005$ moves slightly as $N_t$ increases from 12 to 16.
The rise in the RG-blocked Polyakov loop occurs at the same $\be_F$ where the \Sb order parameters vanish, unlike the $N_f = 8$ panel of \fig{fig:trans}.
Similarly, the $m = 0.005$ eigenvalue density in \fig{fig:m005} shows no sign of chiral symmetry breaking for any $\be_F$; it appears to move straight from the \Sb phase to a chirally symmetric system with $\rho(0) \approx 0$.

How should we interpret our $N_f = 8$ results at $m = 0.005$?
\fig{fig:phases} shows that the 8-flavor finite-temperature transitions occur at steadily stronger couplings as the mass decreases.
At $N_t = 16$, for example, the transition moves from $\be_F = 4.85$ to 4.75 as the mass decreases from $m = 0.015$ to 0.01.
We suspect that at $m = 0.005$, the $N_t = 16$ finite-temperature transition is coincidentally located close enough to the \Sb phase that it becomes entangled with the transition separating the \Sb phase from the weak-coupling phase.
This seems a more modest conjecture than the alternative that the nature of the 8-flavor system changes fundamentally for $m \leq 0.005$ compared to $m \geq 0.01$.

We are currently generating $40^3\X20$ gauge configurations to clarify this situation, which we will discuss in a future publication.
If the $N_t = 16$ finite-temperature transition at $m = 0.005$ is indeed being affected by its proximity to the \Sb phase, then the corresponding $N_t = 20$ transition should be at weak enough coupling to scale in rough agreement with the renormalization group prediction for $\De\be_F$, similarly to $N_t = 16$ at $m \geq 0.01$.
We are also performing $40^3\X20$ calculations for the $N_f = 12$ system, to continue our program of directly comparing these two theories.


\mutesection{Acknowledgments} 
We thank Julius Kuti for helpful comments and discussions, and are grateful to Peter Petreczky, Ron Soltz and the HotQCD Collaboration for allowing us to investigate RG-blocked observables on their gauge configurations.
This research was partially supported by the U.S.~Department of Energy (DOE) through Grant No.~DE-FG02-04ER41290 (A.~C., A.~H.\ and D.~S.) and the DOE Office of Science Graduate Fellowship Program administered by the Oak Ridge Institute for Science and Education under Contract No.~DE-AC05-06OR23100 (G.~P.).
Our code is based in part on the MILC Collaboration's public lattice gauge theory software,\footnote{\texttt{http://www.physics.utah.edu/$\sim$detar/milc/}} and on the PReconditioned Iterative MultiMethod Eigensolver (PRIMME) package~\cite{Stathopoulos:2010}.
Numerical calculations were carried out on the HEP-TH and Janus clusters at the University of Colorado; at Fermilab and Jefferson Lab under the auspices of USQCD supported by the DOE SciDAC program; and at the San Diego Computing Center through XSEDE supported by National Science Foundation Grant No.~OCI-1053575.

\bibliographystyle{utphys}
\bibliography{finiteT}
\end{document}